# A CODER-DECODER MODEL FOR USE
# IN LOSSLESS DATA COMPRESSION

## Abstract

This article describes a technique of using a trigonometric function and combinatorial calculations to code or transform any finite sequence of binary numbers (0s and 1s) of any length to a unique set of three Real numbers.  In reverse, these three Real numbers can be used independently to reconstruct the original Binary sequence precisely.  The main principles of this technique are then applied in a proposal for a highly efficient model for Lossless Data Compression.

## Purpose & Theoretical Background

The core of this technique makes use of the observation that vertical projections of successive circular sector chords can form a sequence of values that are of unique properties and can be mapped exactly to any Binary sequence.  The purpose of this technique is to utilize these properties in order to devise a Coder-Decoder Transform and a potential Data Compression algorithm.

In particular, we will deduce an ordinary Trigonometric function to map a quadrant of a circle to a given Binary sequence of finite length and yield three Real numbers that are unique and can be derived directly from the original sequence of Binary numbers.  These unique three Real numbers can be then used independently to reconstruct the original Binary sequence precisely.

To illustrate how to arrive at this Trigonometric function, consider a circle of radius r having its first quadrant divided into N equal circular sectors.

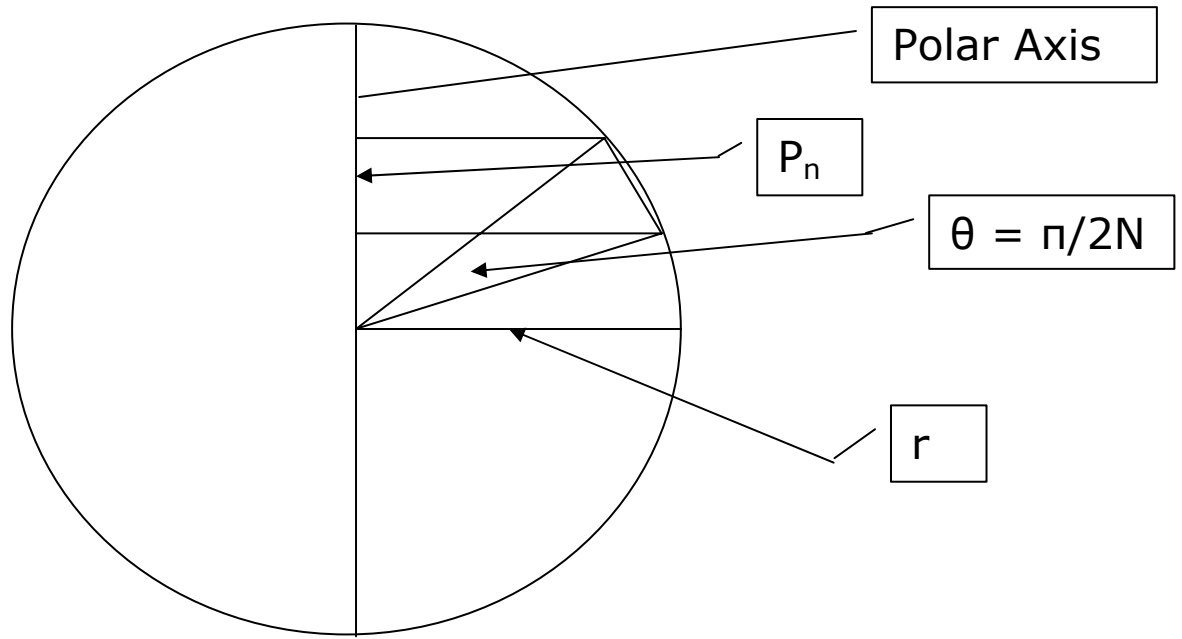

The above diagram depicts one such generic circular sector for the purpose of visualizing this technique.  It follows that the central angle, $\theta$, for each generic sector is $\theta = \pi/2N$

Using the Pythagorean Theorem, the Projection, $P_n$, of the chord of the nth circular sector to the Polar axis is

$$P_n = r (\sin n\theta - \sin (n-1)\theta)$$

Let r = 1, in which case

$$P_n = \sin n\theta - \sin (n-1)\theta$$

or $P_n = \sin (n\pi/2N) - \sin [(n-1)\pi/2N]$

This Trigonometric function is the main Coder Transform that can be used to yield successive values of $P_n$ that are unique and correspond to each individual term of a given Binary sequence exactly.

### Coder-Decoder Concept

We will now demonstrate how a given sequence of Binary numbers can be coded using this function. For this purpose, select a sequence of Binary numbers of **N** terms, i.e.

$$a_1, a_2, a_3, \ldots , a_n, \ldots , a_N$$

where $a_n$ is either a 0 or 1 and the sequence contains a specified number, **Z**, of 1s, i.e. Z is the count of 1s in the selected Binary sequence. A specific and actual numerical example will also be presented.

A. CODING

Numerically assign each term from the selected sequence of Binary numbers that is equal to 1 to a corresponding sequential circular sector and compute the Projection of its chord to the Polar axis for each of these sectors.

As an example, if the $3^{rd}$, $4^{th}$, $8^{th}$, $11^{th}$ terms in the original Binary sequence are 1, then compute the projection of the chord for each of the $3^{rd}$, $4^{th}$, $8^{th}$, $11^{th}$ sectors, i.e. $P_3, P_4, P_8, P_{11}$ using the $P_n$ formula shown above.

Then add all such Projections that correspond to a 1 to yield **R**, i.e. the sum of all Projections for all chords of circular sectors that sequentially correspond to a 1. As a generic example:
$$R = P_i + P_j + P_k + P_l + P_m + \ldots + P_x$$
where the terms $a_i, a_j, a_k, a_l, a_m, \ldots, a_x$ are 1s

Using the example above,
$$R = P_3 + P_4 + P_8 + P_{11}$$

By default, N and Z are positive finite Integers and R has a value between 0 and 1.

The main premise here is that the three Real numbers **N**, **Z**, and **R** constitute a unique combination of values for each selected sequence of Binary numbers since:

- Each $P_n$ is unique

- By default $P_n > P_{n-1} > P_{n-2} > P_{n-3}$ ...

- All values of $P_n$ for a selected sequence of Binary numbers of N terms and having Z number of 1s are taken singularly to compute the sum, R.

B. <u>DECODING</u>

In this section, we will use the three Real numbers (N, Z, R) derived from the previous calculations to reconstruct the original sequence of Binary numbers independently.

To do so, begin with a complete table of all calculated values for $P_n$ from 1 through N, i.e.

$$P_1, P_2, P_3, ... , P_N$$

Then, select all possible combinations of any Z values of $P_n$ from that table. For example if the $P_n$ terms are

$$P_1, P_2, P_3, P_4, P_5 \text{ and } Z = 3$$

Then the combinations are:

$$(P_1, P_2, P_3)$$
$$(P_1, P_2, P_4)$$
$$(P_1, P_2, P_5)$$
$$\cdot$$
$$(P_i, P_j, P_y)$$
$$\cdot$$
$$(P_3, P_4, P_5)$$

Further, compute the sum, $C_n$, for each combination. For the example above, these sums would be:

$$C_1 = P_1 + P_2 + P_3$$
$$C_2 = P_1 + P_2 + P_4$$
$$C_3 = P_1 + P_2 + P_5$$
$$\cdot$$
$$C_n = P_i + P_j + P_y$$
$$\cdot$$
$$C_k = P_3 + P_4 + P_5$$

From Combination Theory, we know that there are k such combinations, where

$$k = N! / [Z! (N-Z)!]$$

As illustrated above, let $C_1, C_2, C_3, ... , C_n, ... , C_k$ be the sums of k combinations for all Z $P_n$'s from that table. Based primarily on the main premise of uniqueness for these terms and as expressed in a previous section, there is a unique $C_x$ in this set where

$$C_x = R$$

and which by default will correspond to a unique sequence of N Binary numbers containing exactly Z terms equal to 1, each of which is situated in the nth position and as determined by each $P_n$ used to compute $C_x$.

As a generic example, if
$$R = C_x = P_3 + P_4 + P_8 + P_{11} + ...$$

then the original sequence is
$$0, 0, 1, 1, 0, 0, 0, 1, 0, 0, 1, ...$$
$$\text{where each nth term} = 1$$

i.e. the subscript of each $P_n$ included in the sum, $C_x$, where $C_x = R$, indicates the position of a 1 in the original Binary sequence.

## A Specific Numerical Example

For this example and in order to demonstrate fully this Coder-Decoder process and see it in action with clear visibility of all steps involved, we will use MS Excel and include all related computations as they appear.

Additionally, because the k value of Z combinations of $P_n$'s can increase dramatically for a large N and will quickly exceed Excel's capacity to handle large decimals, I will limit the number of terms N to 6 and the number Z of 1s to 4.

For this example, let
$$0, 1, 0, 1, 1, 1$$

be the original sequence to code.  In this case,

| Position (n) | Sequence of 0s & 1s | Projection (Pn) | Projection to include |
|---|---|---|---|
| 1 | 0 | 0.258819045 | 0 |
| 2 | 1 | 0.241180955 | 0.241180955 |
| 3 | 0 | 0.207106781 | 0 |
| 4 | 1 | 0.158918623 | 0.158918623 |
| 5 | 1 | 0.099900423 | 0.099900423 |
| 6 | 1 | 0.034074174 | 0.034074174 |
| **SUMS:** | 4 | 1.0000000000 | 0.534074174 = R |

and therefore the original Binary sequence has been coded or transformed to these three Real numbers:

    N = 6
    Z = 4
    R = 0.534074174

***NOTE:***  The excel Table above contains a column titled "Projection to include" which duplicates the values of $P_n$ that correspond to a 1 in the original Binary sequence so that

they can be added to yield R. It is actually a simple excel conditional formula =IF(D7=1,E7,0) meaning "If cell DX = 1, then enter the value from EX cell, otherwise enter 0", where X is the Row Number. In this case, column D is the actual sequence of 0s & 1s and the others follow accordingly. If you wish to try this yourself, and for your convenience, the Excel formula for $P_n$ is =SIN(C7*(PI()/(2*$B$2)))-(SIN((C7-1)*(PI()/(2*$B$2))))

To decode using only this (N, Z, R) set of values and reconstruct the original sequence, we must consider that there are k = 15 possible combinations of sums of any 4 $P_n$'s, and only one of these combinations is equal to R. Therefore:

| Position (n) | Projection (Pn) | Combinations (n of all Z $P_n$'s) | Sum of Combinations (Cx) |
|---|---|---|---|
| 1 | 0.258819045 | 1+2+3+4 | 0.866025404 |
| 2 | 0.241180955 | 1+3+4+5 | 0.724744871 |
| 3 | 0.207106781 | 1+2+5+6 | 0.633974596 |
| 4 | 0.158918623 | 2+3+4+5 | 0.707106781 |
| 5 | 0.099900423 | **2+4+5+6** | **0.534074174 = R** |
| 6 | 0.034074174 | 3+4+5+6 | 0.500000000 |
|   |   | 2+3+5+6 | 0.582262332 |
|   |   | 1+3+4+6 | 0.658918623 |
|   |   | 1+2+4+6 | 0.692992796 |
|   |   | 1+3+5+6 | 0.599900423 |
|   |   | 1+4+5+6 | 0.551712264 |
|   |   | 1+2+3+6 | 0.741180955 |
|   |   | 2+3+4+6 | 0.641280532 |
|   |   | 1+2+3+5 | 0.807007204 |
|   |   | 1+2+4+5 | 0.758819045 |

that means that the original sequence must have a 1 in the $2^{nd}$, $4^{th}$, $5^{th}$, and $6^{th}$ position, i.e.

0, 1, 0, 1, 1, 1

and which is identical to the original Binary sequence selected for this example.

### A Proposal for a Lossless Data Compression Model

For a truly Lossless Compression using this technique exactly as described, the total amount of information in bits required to transmit either the original Binary sequence or its corresponding unique set of three Real numbers (N, Z, R) is the same. This not only follows from basic Information Theory principles, but for this particular method it also becomes apparent quickly once one realizes how large the number of decimal places becomes for the value of R for any set of fairly long binary sequence strings.

However, it is possible to keep at least one of the three Real numbers constant and use pre-populated and pre-stored Key Tables available at both ends of a transmission. This would minimize significantly the required amount of information that needs to be transmitted since only two Integer numbers need to be communicated in order to maintain true lossless compression for each Binary string regardless of its length.

For example, in a generic format such a transmission would require that the number of terms in the original Binary sequence, N, is kept constant. Note that this N number can be selected to be of any size, i.e. in principle, the Binary strings to be transmitted can be as long as desired. Additionally, the required Key Tables, one for each possible value of Z and with all possible Sums of potential Combinations, $C_x$, can be made available at both ends ahead of time and in accordance to how the majority of many popular Compression algorithms work today. Only one Set of Z Key Tables needs to be made available to enable this method.

In this case and using the same symbols as before, we already know **N** since we select it as a constant and that $0 \geq Z \geq N$. Then it follows that there are

$$K = \sum_{z=0}^{Z} k_z$$

where $k_z = N! / [Z! (N-Z)!]$

possible values of R for each value of Z. As it is also known, in general this Sum **K** is actually reduced to $X^Z$ where X is the number of different digits in the sequence, i.e. for our case of Binary strings $X = 2$ and $X^Z = 2^Z$.

In their most simple and intuitive format, these Key Tables would each consist of two columns, one having sequentially increasing integer numbers of 1 through K and the other all potential values of R for a given value of Z. Only one of these rows, $(k_x, R_x)$, is such that $R_x = R$ for any given Binary sequence. As a result, only the integers Z and $k_x$ need to be transmitted as a coded representation of the original Binary sequence.

Also, since the highest possible value for Z is N in any given Binary sequence, the maximum number of Key Tables required is N and each has a maximum of $2^Z$ rows. In other words, the two Integers to be transmitted for each string are never larger than N or $2^Z$. This also offers some additional perspective on the size of the corresponding Key Tables, i.e. they are relatively small.

In essence, this method can be adapted to using a simple pre-populated database at either end of a transmission that enables much faster processing in an Indexed Retrieval "look-up" manner. Naturally, the efficiency of this method increases exponentially with longer Binary strings and the corresponding database can be formatted in various clever ways to enhance retrieval efficiency even further.

Furthermore and in most real-world applications, many Binary strings contained in a typical Data Transmission are repeated profusely. Although this repeat frequency may theoretically suffer with extremely large values of N, in most practical applications the $(Z, k_x)$ Integer pairs appear numerous times. Therefore, a potential improvement would be to group all $(Z, k_x)$ Integer pairs and transmit them as one individual pair along with some simple Frequency or Positional qualifiers. These qualifiers then can be used at the receiving end to re-assemble fairly large strings efficiently using already developed techniques and in a similar fashion to most Compression systems in use today.

Please note that this method does not depend on any Character Sequential Preferences or predetermined Frequency or Syntactical Occurrence Values that most current systems rely upon, especially for text transmissions in English or most other common languages. It is totally independent of Language or Character Set and is entirely Object Oriented as long as these Objects can be expressed in finite Binary strings of any length.

For this reason, objects such as images or video and similar non-text files would benefit the most in terms of enhanced Lossless Compression and efficient transmission using this method.

## Summary

This article expands on a technique for a Coder-Decoder or mathematical Transform that is based on mapping a finite Binary string of any length onto vertical projections of circular sector chords, which in turn yields three Real numbers that are unique and reciprocal to the original Binary string. Examples of how this technique can be used are cited fully and in detail. It is also shown that all calculations required to use this technique are based solely on numerical and combinatorial computations that are derived from or can be applied directly on any Binary string to be coded and are amenable to using a Computer system.

It is also described how the main principles of this technique can be expanded in creating an efficient model for a Lossless Data Compression algorithm. It further suggests the use of well tested methods of using data processing or database retrieval formats to enhance the efficiency of this proposal.

Implementation of this model can be readily applied to any finite Binary string and is completely Object Oriented. It is independent of syntactical or character redundancy schemes often used in Compression of text. For this reason, it can maintain its efficacy as Lossless on Images or Video and similar files.

By: Alex Papalexis
 Sunnyvale, CA
Cell: (408) 390-9000